\documentclass[authoryear,final,1p,times]{elsarticle}

\usepackage{amssymb}
\usepackage{amsmath}
\usepackage{booktabs}
\usepackage{longtable}
\usepackage{siunitx}
\usepackage{multirow}
\usepackage{url}

\begin{document}

\begin{frontmatter}

\title{Distinguishing case-mix from context heterogeneity in prognostic regression model synthesis settings}

\author[imbi,fdm]{Max Behrens}
\author[nephro]{Janis M. Nolde}
\author[pneumo]{Eleni Papakonstantinou}
\author[mathstoch]{Gabriele Bellerino}
\author[imbi]{Theodoros Evrenoglou}
\author[mathstoch,fdm]{Angelika Rohde}
\author[pneumo]{Daiana Stolz}
\author[imbi,fdm]{Moritz Hess}
\author[imbi,fdm]{Harald Binder}

\affiliation[imbi]{organization={Institute of Medical Biometry and Statistics, Faculty of Medicine and Medical Center, University of Freiburg},
            addressline={Stefan-Meier-Strasse 26}, 
            city={Freiburg},
            postcode={79104}, 
            country={Germany}}

\affiliation[fdm]{organization={Freiburg Center for Data Analysis, Modeling and AI, University of Freiburg},
            addressline={Ernst-Zermelo-Strasse 1}, 
            city={Freiburg},
            postcode={79104}, 
            country={Germany}}

\affiliation[nephro]{organization={Department of Nephrology, Faculty of Medicine and Medical Center, University of Freiburg},
            addressline={Hugstetter Strasse 55}, 
            city={Freiburg},
            postcode={79106}, 
            country={Germany}}

\affiliation[pneumo]{organization={Clinic of Pneumology, Medical Center – University of Freiburg, Faculty of Medicine, University of Freiburg},
            addressline={Killianstrasse 5}, 
            city={Freiburg},
            postcode={79106}, 
            country={Germany}}

\affiliation[mathstoch]{organization={Department of Mathematical Stochastics, University of Freiburg},
            addressline={Ernst-Zermelo-Strasse 1}, 
            city={Freiburg},
            postcode={79104}, 
            country={Germany}}

\begin{abstract}
Prognostic regression models often synthesize data from multiple sites, whether within a multi-site study, across federated settings, or in individual participant data meta-analysis. Here, a site is any data source, such as a hospital, registry, trial, or study, and need not be a physical center. Analysts must then decide whether one regression model represents all sites or whether site-specific models are needed. Established measures such as coefficient-level $\tau^2$ quantify heterogeneity but do not distinguish its source. We focus on diagnosing whether coefficient heterogeneity reflects case-mix or site-specific context effects. Case-mix heterogeneity can arise when linear regression terms approximate multivariable non-linear relationships in populations with different covariate distributions. Contextual heterogeneity arises when comparable patients require different regression relationships across sites. We do this by fitting site-specific local regressions in a dimension-reduced space and partitioning the smoothed coefficient surfaces into a cross-site reference and site-specific deviations. An autoencoder and custom loss structure the latent space around local prognostic relationships. We then project this partition onto the outcome scale to derive observation- and site-level summaries. We demonstrate the approach on a COPD trial with two sites. In the three leading latent slope coordinates, coefficient-surface variation was predominantly contextual. The derived observation-level outcome-scale variance partition was case-mix-leading, whereas its between-site aggregation was concentrated in contextual differences rather than case-mix shifts. A permuted-site negative control assesses whether the contextual summary can arise when site labels carry no signal. This diagnostic distinction can inform whether joint or site-specific regression models should be evaluated.
\end{abstract}

%% Keywords
\begin{keyword}
evidence synthesis \sep regression models \sep individual patient data \sep heterogeneity \sep autoencoder \sep local regression

\end{keyword}

\end{frontmatter}

%% main text 

\section{Introduction}

Meta-analyses of clinical prognostic models often integrate evidence from sources for which individual patient data (IPD) are available with aggregate results from others \citep{debrayIndividualParticipantData2015, rileyMetaanalysisIndividualParticipant2010a}. We use \emph{site} for any such data source, including a hospital, registry, clinical trial, or study; it need not be a physical center. This mixed-data context forces a synthesis decision: whether to pool prognostic associations into a single summary model or to model site-specific estimates separately. Pooling under substantial heterogeneity can produce misleading estimates \citep{panagiotopoulouMetaAnalysisModels2025}, whereas unnecessary stratification reduces precision.

The appropriateness of pooling depends on the source of heterogeneity. Consider two sites relating a common predictor set to the same outcome. If they sample different regions of the predictor distribution and the outcome relationship is nonlinear or modified by patient characteristics, their average coefficients can differ even when the conditional relationship is shared \citep{rileyIndividualParticipantData2020}. This is case-mix heterogeneity: disagreement caused by who is observed. If instead comparable patients require different regression relationships at the two sites, the disagreement is contextual, arising from where or how the data were generated.

These two sources of heterogeneity have different implications for evidence synthesis \citep{voNovelApproachIdentifying2019, wangCausalMetaR2025}. Case-mix heterogeneity includes effect modification by patient-level covariates and non-linear effects observed at different covariate ranges \citep{rileyIndividualParticipantData2020}. When the conditional relationship is shared, it can be addressed by modeling that structure and standardizing to a relevant covariate distribution, so pooling may be appropriate under adequate adjustment, overlap, and functional-form specification. Contextual heterogeneity instead arises from effect modification by site-level factors such as treatment protocols, measurement standards, or care infrastructure, and can make pooled estimates misleading even after adjustment. The distinction remains conditional on the measured predictors: unmeasured patient characteristics that vary systematically across sites may appear as contextual effects \citep{dahabrehUsingGroupData2016}.

Existing meta-analytic methods can flag disagreement but do not identify the channel through which it arises. A coefficient-level $\tau^2$ warns that site-specific estimates differ without saying whether the difference reflects who was enrolled, how comparable patients map to the outcome, or both. One-stage IPD meta-analysis, such as mixed-effects models with random intercepts and slopes, captures between-study coefficient variation but does not separate case-mix from contextual components \citep{burkeMetaanalysisUsingIndividual2017, debrayIndividualParticipantData2015}, and explicit predictor-by-site interactions become impractical as the number of predictors grows. Meta-regression \citep{thompsonHowShouldMetaregression2002} attributes heterogeneity to study-level moderators, but requires them to be pre-specified; unknown or unmeasured moderators leave it unexplained. Location-scale models let heterogeneity depend on study-level predictors \citep{viechtbauerLocationScaleModels2022}, but typically assume a known and simple structure. None of these yields the diagnostic information needed to choose between pooling and site-specific synthesis.

We propose using sites with available IPD as a diagnostic sample to assess whether coefficient heterogeneity reflects case-mix, context, or their covariation. Building on our earlier comparison of global and patient-specific regression relationships \citep{behrensContrastingGlobalPatientSpecific2026}, we use local models to attribute prognostic variation to position-driven and site-specific sources. The primary object is a partition of site-specific local coefficient surfaces into a cross-site reference and site-specific deviations. These coefficients are defined with respect to learned latent coordinates rather than the original predictors, but the partition retains the coefficient-level structure familiar from meta-analysis. We then project this partition onto the outcome scale to obtain scalar observation- and site-level summaries. This source-attribution diagnostic can inform whether pooled or site-specific synthesis should be evaluated.

Local regression provides the structure needed for this source attribution. It captures effect modification and non-linearity across patient subgroups without requiring those subgroups to be defined in advance \citep{tutzLocalizedClassification2005}, so comparing site-specific local models at matched patient profiles separates position-driven variation from differences between sites. With many prognostic factors, however, local regression is infeasible in the original variable space \citep{bellmanApproximationCurvesLine1961, loaderOriginsLocalRegression1999}. We therefore learn a low-dimensional representation using an autoencoder \citep{hintonReducingDimensionalityData2006a} adapted through outcome-based losses, so that proximity reflects similarity in prognostic relationships \citep[e.g.][]{Hackenberg2025DomainAdaptation, behrensContrastingGlobalPatientSpecific2026}.

This strategy complements approaches that fit subgroup models using supervised machine learning, such as tree-based subgroup identification \citep{huberMetaMOB2022}, or use unsupervised deep neural networks to study prognostic heterogeneity \citep{westUnsupervisedDL2025}. The framework requires IPD, but may still inform mixed-data synthesis when only some sites provide individual-level data. Such use depends on whether the heterogeneity pattern observed at the IPD sites generalizes to the remaining evidence.

Section~\ref{sec:methods} defines the coefficient-surface partition, its outcome-scale summaries, and how the latent representation is learned. Section~\ref{sec:results} illustrates the method in a multi-site COPD dataset with over 30 prognostic factors, with a model-swap diagnostic and permuted-site negative control. Section~\ref{sec:discussion} covers the generalizability assumption, implications for meta-analysis, and potential extensions.

\section{Methods} \label{sec:methods}

\subsection{Framework Overview and Notation}

The method is a diagnostic for multi-site prognostic modeling, not a new global prediction model. It has three conceptual steps. First, a low-dimensional representation defines observation profiles at which local models can be compared. Second, site-specific local regressions estimate a coefficient surface over those profiles. Third, each site surface is partitioned into a cross-site reference and a site-specific departure. This coefficient-surface partition is the primary analysis. Projection onto the linear-predictor scale and aggregation by site provide two derived outcome-scale summaries.

Formally, consider $n$ observations from $C$ sites, each with a common set of $p$ predictors $\mathbf{x}_i \in \mathbb{R}^p$, outcome $y_i$, and site index $c_i$ for $i = 1, \dots, n$. Each observation denotes one eligible analysis record. Stacking the predictor vectors gives $\mathbf{X} \in \mathbb{R}^{n \times p}$, and $n_k$ denotes the number of observations in site $k$. A learned mapping represents $\mathbf{x}_i$ by $\mathbf{z}_i \in \mathbb{R}^D$, with $D \ll p$, and applying it to all observations gives $\mathbf{Z} \in \mathbb{R}^{n \times D}$; the $r$-th coordinate of $\mathbf{z}_i$ is written $z_{ir}$. Here, $c_i$ denotes the site membership of observation $i$, whereas $k \in \{1,\dots,C\}$ serves as a running site index. Latent coordinates are indexed by $r=0,\dots,D-1$, matching the zero-based labels used in the figures.

The low-dimensional representation is needed because local regression becomes unstable when neighborhoods are defined directly in a high-dimensional predictor space. We first define the local models and the diagnostic conditional on $\mathbf{Z}$. Section~\ref{sec:composite_loss} then explains how an autoencoder learns $\mathbf{Z}$ jointly with the outcome-based objectives.

\subsection{Site-Specific Local Regression in Latent Space}

Given $\mathbf{Z}$, site-specific local generalized linear models (GLMs) describe how prognostic relationships vary across latent observation profiles. Within a sufficiently small neighborhood, the relationship between latent coordinates and the transformed outcome mean is approximated by a linear predictor, even when the global relationship is nonlinear. A separate local model is therefore estimated for each site $k$ at each query position $\mathbf{z}_i$.

For each observation $i$ and site $k \in \{1, \dots, C\}$, the general local model can be written as a weighted GLM,
\[
g\!\left(\mu_{ij}^{(k)}\right) = \alpha_i^{(k)} + {\boldsymbol{\beta}_i^{(k)}}^T \mathbf{z}_j,
\]
where $g(\cdot)$ is the link function, $\mu_{ij}^{(k)}$ is the expected outcome for observation $j$ under the local model centered at $\mathbf{z}_i$ in site $k$, and $\boldsymbol{\beta}_i^{(k)} \in \mathbb{R}^{D}$ is the local slope vector. Thus, $i$ indexes the query position, $j$ indexes observations used in the local fit, and $k$ indexes the site-specific model. This framing provides a route to other outcome types through the likelihood and link function. The present COPD application uses a continuous outcome with a Gaussian identity-link working model, so estimation reduces to ridge-stabilized weighted least squares:
\[
\begin{aligned}
(\hat{\alpha}_i^{(k)}, \hat{\boldsymbol{\beta}}_i^{(k)}) = \arg\min_{\alpha, \boldsymbol{\beta}} \bigg[
&\sum_{j: c_j = k} w_{ij}^{(k)} \left(y_j - \alpha - \boldsymbol{\beta}^T \mathbf{z}_j\right)^2 \\
&+ \lambda_{\mathrm{ridge}}\left(\alpha^2 + \|\boldsymbol{\beta}\|_2^2\right)
\bigg],
\end{aligned}
\]
where $w_{ij}^{(k)}$ are site-specific kernel weights that assign greater influence to observations $j$ closer to the query point $\mathbf{z}_i$, and $\lambda_{\mathrm{ridge}}=10^{-4}$ is a small stabilizer for near-singular local fits. These models relate outcomes to latent coordinates. The outcome-scale summaries below use the identity-link case; extensions to non-identity links would require specifying the corresponding outcome-scale transformation.

The weights $w_{ij}^{(k)}$ are defined via a Gaussian kernel on normalized distances:
\[
w_{ij}^{(k)} = K(u_{ij}^{(k)}), \quad u_{ij}^{(k)} = \frac{\|\mathbf{z}_i - \mathbf{z}_j\|_2}{b_i^{(k)}}, \quad K(u) = \exp\left(-\frac{u^2}{2\sigma^2}\right),
\]
where $b_i^{(k)}$ is an adaptive site-specific bandwidth and $\sigma$ controls how sharply weights decay. Let $n_{k,\mathrm{ref}}$ denote the number of observations from site $k$ in the reference set used for the current local fit. In the implementation, these reference-set distances for site $k$ are sorted, and $b_i^{(k)}$ is selected at zero-based index $\lfloor k_{\mathrm{NN}}n_{k,\mathrm{ref}}\rfloor$, corresponding to order-statistic rank $\lfloor k_{\mathrm{NN}}n_{k,\mathrm{ref}}\rfloor+1$. The neighborhood fraction therefore determines $b_i^{(k)}$, while the effective distance bandwidth is $\sigma b_i^{(k)}$. Smaller values yield more local and flexible estimates, whereas larger values yield smoother estimates. Tuning details are provided in the Appendix.

Four roles for weights should be distinguished. The kernel weights $w_{ij}^{(k)}$ determine which nearby observations influence a local regression. Sampling or design weights determine how sites contribute observations during optimization; the COPD implementation used balanced mini-batches. Reference weights $q_k$ define the cross-site coefficient surface in Section~\ref{sec:variance_decomp}; the application uses a reference assigning equal weight to each site, $q_k=1/C$. Aggregation weights $\pi_k$ define the target population for the site-level summary in Section~\ref{sec:outcome_decomp}; the application uses the empirical shares $n_k/n$, while alternative prespecified shares could represent another target population. The laws of total variance and covariance in Section~\ref{sec:outcome_decomp} are exact only for $\pi_k=n_k/n$; under prespecified shares $\tau_M^2$ remains interpretable as a between-site summary, but it is no longer a component of $\mathrm{Var}_i(M_i)$. Neither $q_k$ nor $\pi_k$ is a latent-distance weight.

The pointwise local-WLS estimates are smoothed component-wise to obtain a fitted intercept surface $\alpha^{(k)}(\mathbf{z})$ and slope surface $\boldsymbol{\beta}^{(k)}(\mathbf{z})\in\mathbb{R}^D$ for each site. For each component, KernelRidge regression with an RBF kernel is fitted to the stacked local-WLS estimates at the observed latent positions from that site, requiring at least 10 observations. The fitted surfaces are evaluated at every observed latent position. Hats are suppressed after smoothing. The resulting partition is therefore a fitted, sample-level diagnostic rather than a population-identified decomposition.

Reliable estimation requires sufficient local sample sizes and common support. Each site must contribute enough observations in shared regions of the latent space for its fitted surfaces to be compared at the same positions. If sites occupy disjoint regions, contextual departures cannot be distinguished from extrapolation. Support can be assessed in the original covariate space, for example through propensity-score distributions for site membership, and along the learned coordinates using site-specific density summaries; Figure~\ref{fig:latent_dimension_sources}B shows these summaries for the three leading contextual dimensions.

\subsection{Coefficient-Surface Partition} \label{sec:variance_decomp}

At each latent profile $\mathbf{z}$, the fitted intercept $\alpha^{(k)}(\mathbf{z})$ and slope vector $\boldsymbol{\beta}^{(k)}(\mathbf{z})$ describe the local prognostic relationship in site $k$. We compare the slope surfaces with a prespecified weighted cross-site reference. The resulting identity separates position-driven variation in the reference surface from site-specific departures evaluated at the same positions. This coefficient-surface partition is the primary analysis. Outcome-scale projection and site aggregation are introduced separately in Section~\ref{sec:outcome_decomp}.

Using reference weights $q_k$ with $\sum_k q_k=1$, define the reference slope surface and the site-specific departure as
\begin{align}
  \bar{\boldsymbol{\beta}}(\mathbf{z})
  &= \sum_{k=1}^C q_k\boldsymbol{\beta}^{(k)}(\mathbf{z}),
  \qquad q_k=\frac{1}{C}\ \text{in the application}, \\
  \boldsymbol{\beta}^{(k),\mathrm{ctx}}(\mathbf{z})
  &= \boldsymbol{\beta}^{(k)}(\mathbf{z})-\bar{\boldsymbol{\beta}}(\mathbf{z}).
\end{align}
For observation $i$ from site $c_i$, the slope partition is therefore
\[
  \boldsymbol{\beta}^{(c_i)}(\mathbf{z}_i)
  = \bar{\boldsymbol{\beta}}(\mathbf{z}_i)
    +\boldsymbol{\beta}^{(c_i),\mathrm{ctx}}(\mathbf{z}_i).
\]
The intercept is partitioned analogously, with $\bar{\alpha}(\mathbf{z})=\sum_{k=1}^Cq_k\alpha^{(k)}(\mathbf{z})$ and $\alpha^{(k),\mathrm{ctx}}(\mathbf{z})=\alpha^{(k)}(\mathbf{z})-\bar{\alpha}(\mathbf{z})$. Hence, $\alpha^{(c_i)}(\mathbf{z}_i)=\bar{\alpha}(\mathbf{z}_i)+\alpha^{(c_i),\mathrm{ctx}}(\mathbf{z}_i)$.
The pointwise local-WLS estimates fluctuate around these fitted smoothed surfaces. The reported partition uses the smoothed intercepts and slopes and excludes the remaining pointwise fitting variation.

Let $\mathrm{Var}_i$ and $\mathrm{Cov}_i$ denote empirical variance and covariance over the $n$ observed profiles, using divisor $n$; the reported percentage shares are unchanged if divisor $n-1$ is used consistently. For slope coordinate $r=0,\ldots,D-1$, the coefficient-space variance identity is
\[
  \mathrm{Var}_i\!\left\{\beta_r^{(c_i)}(\mathbf{z}_i)\right\}
  = \mathrm{Var}_i\!\left\{\bar{\beta}_{r}(\mathbf{z}_i)\right\}
  + \mathrm{Var}_i\!\left\{\beta_r^{(c_i),\mathrm{ctx}}(\mathbf{z}_i)\right\}
  + 2\,\mathrm{Cov}_i\!\left\{\bar{\beta}_{r}(\mathbf{z}_i),
      \beta_r^{(c_i),\mathrm{ctx}}(\mathbf{z}_i)\right\}.
\]
This identity is the primary decomposition; the intercept follows the same form with $\alpha$ in place of $\beta_r$. Figure~\ref{fig:latent_dimension_sources}A applies the slope identity to selected latent coordinates. Its first term is position-driven reference-surface variation. It becomes between-site case-mix heterogeneity only when the distributions of latent positions, and therefore site summaries of this term, differ between sites. The second term is variation in site-specific departures at matched positions. The covariance need not vanish even though $\sum_k q_k\boldsymbol{\beta}^{(k),\mathrm{ctx}}(\mathbf{z})=\mathbf{0}$ at each fixed $\mathbf{z}$: the empirical identity evaluates only each observation's own-site departure over the observed, site-specific position distributions.

Because these coordinates belong to a learned latent representation, their slopes and coordinate-wise shares are conditional on the fitted scale and orientation. They are not original-variable regression coefficients and cannot be compared dimension by dimension across independently trained encoders without alignment.

\subsection{Derived Outcome-Scale Summaries} \label{sec:outcome_decomp}

The first derived summary projects the coefficient-surface partition onto the linear-predictor scale. For observation $i$, define
\begin{align}
  M_i^{\mathrm{mix}}
  &= \bar{\alpha}(\mathbf{z}_i)
     +\mathbf{z}_i^T\bar{\boldsymbol{\beta}}(\mathbf{z}_i), \\
  M_i^{\mathrm{ctx}}
  &= \alpha^{(c_i),\mathrm{ctx}}(\mathbf{z}_i)
     +\mathbf{z}_i^T\boldsymbol{\beta}^{(c_i),\mathrm{ctx}}(\mathbf{z}_i), \\
  M_i
  &= M_i^{\mathrm{mix}}+M_i^{\mathrm{ctx}}
   = \alpha^{(c_i)}(\mathbf{z}_i)
     +\mathbf{z}_i^T\boldsymbol{\beta}^{(c_i)}(\mathbf{z}_i).
\end{align}
Here, $M_i^{\mathrm{mix}}$ is the position-driven outcome component for observation $i$ under the cross-site reference relationship. We label it the observation-level case-mix component, but it is not a distributional contrast by itself. Between-site case-mix heterogeneity arises only when the distribution or site mean of this component differs between sites. $M_i^{\mathrm{ctx}}$ is the context component: the site-specific shift at the same observation position. The reported outcome-scale partition uses the smoothed identity $M_i = M_i^{\mathrm{mix}} + M_i^{\mathrm{ctx}}$ and therefore excludes the remaining pointwise local-WLS fitting variation.

The variance of this smoothed outcome-scale component gives the outcome-scale variance partition
\[
  \mathrm{Var}_i(M_i) = \mathrm{Var}_i(M_i^{\mathrm{mix}}) + \mathrm{Var}_i(M_i^{\mathrm{ctx}}) + 2\,\mathrm{Cov}_i(M_i^{\mathrm{mix}}, M_i^{\mathrm{ctx}}).
\]
Thus, $\mathrm{Var}_i(M_i^{\mathrm{mix}})$ is the observation-level position-driven variance, labelled case-mix in this derived summary; $\mathrm{Var}_i(M_i^{\mathrm{ctx}})$ is the variance attributed to context; and $2\,\mathrm{Cov}_i(M_i^{\mathrm{mix}}, M_i^{\mathrm{ctx}})$ measures whether the components reinforce or offset each other. The variance identity is algebraically exact. At least two sites are required for a non-trivial contextual contrast.

Finally, the outcome-scale quantities can be aggregated from observations to sites. Let $n_k$ be the number of observations in site $k$, $\pi_k=n_k/n$, and define the empirical site means
\begin{align}
  \bar{M}_k^{\mathrm{mix}}
  &= \frac{1}{n_k}\sum_{i:c_i=k}M_i^{\mathrm{mix}}, \\
  \bar{M}_k^{\mathrm{ctx}}
  &= \frac{1}{n_k}\sum_{i:c_i=k}M_i^{\mathrm{ctx}}, \\
  \bar{M}_k
  &= \frac{1}{n_k}\sum_{i:c_i=k}M_i
   = \bar{M}_k^{\mathrm{mix}}+\bar{M}_k^{\mathrm{ctx}}.
\end{align}
A descriptive between-site variance for this outcome-scale surface is
\[
  \tau_M^2 = \sum_{k=1}^{C} \pi_k(\bar{M}_k - \bar{M})^2,
\]
where $\bar{M} = \sum_k \pi_k \bar{M}_k$. The coefficient-surface partition is structurally closest to the coefficient heterogeneity familiar from meta-analysis. In contrast, $\tau_M^2$ is a further, population-weighted between-site aggregation of the derived outcome-scale surface. It is therefore a descriptive analogue, not a DerSimonian--Laird estimator of coefficient heterogeneity \citep{dersimonianMetaanalysisClinicalTrials1986}. It is the $n_k/n$-weighted between-site component of $\mathrm{Var}_i(M_i)$. For $C = 2$, it reduces to $\pi_1\pi_2(\bar{M}_1 - \bar{M}_2)^2$. The law of total variance links the observation-level and site-level views:
\[
  \mathrm{Var}_i(A_i)
  = \sum_{k=1}^{C} \pi_k\,\mathrm{Var}(A_i \mid c_i = k)
  + \sum_{k=1}^{C} \pi_k(\bar{A}_k - \bar{A})^2,
  \qquad A_i \in \{M_i^{\mathrm{mix}}, M_i^{\mathrm{ctx}}, M_i\}.
\]
Here $A_i$ and $B_i$ are placeholders for the outcome-scale quantities named at each display, $\bar{A}_k$ and $\bar{B}_k$ denote empirical site means, $\bar{A}=\sum_k\pi_k\bar{A}_k$ and $\bar{B}=\sum_k\pi_k\bar{B}_k$ denote their weighted overall means, and the conditional empirical variance and covariance operators use divisor $n_k$. The covariation component follows the corresponding law of total covariance:
\[
\begin{aligned}
  \mathrm{Cov}_i(A_i,B_i)
  ={}& \sum_{k=1}^{C}\pi_k\,\mathrm{Cov}(A_i,B_i\mid c_i=k) \\
  &+ \sum_{k=1}^{C}\pi_k(\bar{A}_k-\bar{A})(\bar{B}_k-\bar{B}),
  \qquad
  (A_i,B_i)=(M_i^{\mathrm{mix}},M_i^{\mathrm{ctx}}).
\end{aligned}
\]
For $M_i^{\mathrm{mix}}$, the within-site term reflects observation-position spread under the reference relationship, and the between-site term reflects differences in site centroids. For $M_i^{\mathrm{ctx}}$, the between-site term captures the average contextual shift. The within-site term captures how that shift varies across observation positions within a site. Thus, $\tau_M^2$ is the between-site aggregation of the outcome-scale variance partition, not a third decomposition level. In the application, $M_i^{\mathrm{mix}}$, $M_i^{\mathrm{ctx}}$, and $M_i$ are on the standardized SGRQ linear-predictor scale.

\subsection{Learning the Latent Representation} \label{sec:composite_loss}

A single autoencoder is trained on observations pooled across all $C$ sites. The encoder maps $\mathbf{x}_i$ to $\mathbf{z}_i = \operatorname{enc}(\mathbf{x}_i; \boldsymbol{\phi}_{\operatorname{enc}})$, and the decoder reconstructs the input as $\hat{\mathbf{x}}_i = \operatorname{dec}(\mathbf{z}_i; \boldsymbol{\phi}_{\operatorname{dec}})$, with $\hat{\mathbf{X}} \in \mathbb{R}^{n \times p}$ collecting the reconstructions. The network parameters are $\boldsymbol{\phi} = \{\boldsymbol{\phi}_{\operatorname{enc}}, \boldsymbol{\phi}_{\operatorname{dec}}\}$. Unlike a fixed preprocessing step, the representation is learned jointly with outcome-based objectives so that latent proximity supports the local models defined above. Because the encoder is differentiable, its original-predictor contributions to each latent coordinate can subsequently be summarized descriptively using SHAP (Section~\ref{sec:results}).

The objective function has three goals: preserve information from the original predictors, form prognostically meaningful local neighborhoods, and retain site-specific outcome relationships.

The first term ensures $\mathbf{Z}$ retains sufficient information to reconstruct the original predictors:
\[
\text{Loss}_{\text{rec}}(\boldsymbol{\phi}) = \frac{1}{n \cdot p} \| \mathbf{X} - \hat{\mathbf{X}} \|_F^2,
\]
where $\| \cdot \|_F^2$ denotes the squared Frobenius norm. With linear encoder and decoder mappings and squared-error reconstruction, an optimum spans the same $D$-dimensional principal subspace as PCA, although the learned coordinates need not equal the PCA scores. The nonlinear encoder used here can capture more complex structure.

The second term, the local prognostic loss ($\text{Loss}_{\text{local}}$), rewards neighborhoods in which local regression explains outcomes better than an intercept-only model. For each query observation $i$ in site $c_i$,
\[
\text{Loss}_{\text{local}}(\boldsymbol{\phi}) = \frac{1}{n}\sum_{i=1}^{n} -\log\left(\frac{L_i^{(c_i)}}{L_{i,\text{null}}}\right),
\]
where $L_i^{(c_i)}$ and $L_{i,\text{null}}$ are weighted local likelihoods over the kernel-weighted neighborhood centered at $\mathbf{z}_i$. Lower values favor latent neighborhoods whose outcomes follow a shared local prognostic relationship.

The third term, the native-site loss ($\text{Loss}_{\text{native}}$), retains site-specific prognostic differences. It compares each query observation's fit under its own site model with its fit under the other site models:
\[
\text{Loss}_{\text{native}}(\boldsymbol{\phi}) = \frac{1}{n}\sum_{i=1}^{n} \frac{1}{C-1} \sum_{k \neq c_i} -\log\left(\frac{L_i^{(c_i)}}{L_i^{(k)}}\right),
\]
where $L_i^{(k)}$ is the weighted local likelihood under site $k$'s local model. Lower values indicate better fit under the observation's own site model relative to the alternatives.

The final objective combines these components:
\[
\text{Loss}(\boldsymbol{\phi}) = \lambda_{\text{rec}} \text{Loss}_{\text{rec}}(\boldsymbol{\phi}) + \lambda_{\text{local}} \text{Loss}_{\text{local}}(\boldsymbol{\phi}) + \lambda_{\text{native}} \text{Loss}_{\text{native}}(\boldsymbol{\phi}).
\]
The tuning weights balance information retention, prognostically coherent neighborhoods, and site-specific outcome structure. In the identity-link model, the local intercepts and slopes are computed in closed form by ridge-stabilized weighted least squares for the current $\mathbf{Z}$ rather than optimized as free network parameters. Gradients pass through this computation to update $\boldsymbol{\phi}$.

To limit over-separation by site labels, the native-site loss is down-weighted in the earliest encoder layer. The setting is reported in the Appendix, under \emph{Hyperparameter Configuration} and assessed descriptively using the permuted-site negative control.

\subsection{Model Training}

The parameters $\boldsymbol{\phi}$ are estimated by gradient-based minimization of the composite objective. End-to-end optimization is feasible because the computation is differentiable almost everywhere. The encoder is smooth in its parameters, the local weights are smooth between nearest-neighbor rank changes, and the ridge-stabilized weighted least squares solution is differentiable wherever the regularized normal equations are nonsingular.

Training proceeds in random batches. For each batch, observations are encoded, local weighted regressions are recomputed at the current latent positions, and the composite loss updates the encoder and decoder through this computation. Over a full epoch, every observation serves as a query point once. Optimization settings are reported in the Appendix, under \emph{Hyperparameter Configuration}.

\section{Results} \label{sec:results}

\subsection{Clinical Data from COPD Patients}

To evaluate our approach in a realistic setting characterized by potential between-site variability, data from the PREVENT study \citep{stolzIntensifiedTherapyInhaled2018}, a clinical trial with two participating centers, were utilized. We treated each center as a site to illustrate source attribution for coefficient heterogeneity. Two sites are the minimum needed to form the between-site contrast, making this a compact illustration while limiting assessment of multi-site variation. The trial investigated whether intensified inhaled therapy (LABA/ICS) could reduce exacerbations in patients with moderate to severe chronic obstructive pulmonary disease (COPD). The analysis comprised 300 eligible observations ($n_1 = 200$ in Site~1, $n_2 = 100$ in Site~2) after excluding upper-respiratory-tract-infection observations and applying the original per-feature filtering. Predictors were used to predict subsequent SGRQ total; higher SGRQ scores indicate greater disease impact. Thirty-eight lung-function and physiological measurements served as predictors (see Appendix for variable definitions).

All analyses start from the same fixed processed dataset used by the original PREVENT analysis. Its upstream k-nearest-neighbor imputation included the contemporaneous SGRQ total, and correlation pruning and outlier filtering were performed once on the full dataset. The production coefficient-surface partition and all three figures use the standardized version of this fixed input. The observation-level held-out comparison uses the corresponding unscaled version and fits scaling within each training fold, so its RMSE remains on the SGRQ-point scale. Consequently, both the source-attribution and validation estimates are conditional on the fixed upstream preprocessing.

The production model was trained using the Adam optimizer \citep{kingmaAdamMethodStochastic2017} with a learning rate of $10^{-5}$ and repeated across 10 random seeds to assess initialization sensitivity. The two latent-space figures use representative seed 1213, whose outcome-scale component shares were closest to the cross-seed center; results explicitly distinguish representative-seed bootstrap summaries from cross-seed ranges. The observation-level held-out comparison used three seeds per fold. Architecture details, including layer dimensions and activation functions, as well as hyperparameter selection ($\lambda_{\text{rec}}$, $\lambda_{\text{local}}$, $\lambda_{\text{native}}$, neighborhood fraction $k_{\mathrm{NN}}$, kernel scale $\sigma$), are provided in the Appendix. The model was implemented in Python 3.12 using PyTorch 2.2 \citep{paszkePyTorchImperativeStyle2019}, and the code is available at \url{https://github.com/maxjonasbehrens/case-mix-context-decomposition}.

\subsection{Heterogeneity in Original Variable Space}

\begin{figure}
    \centering
    \includegraphics[width=\linewidth]{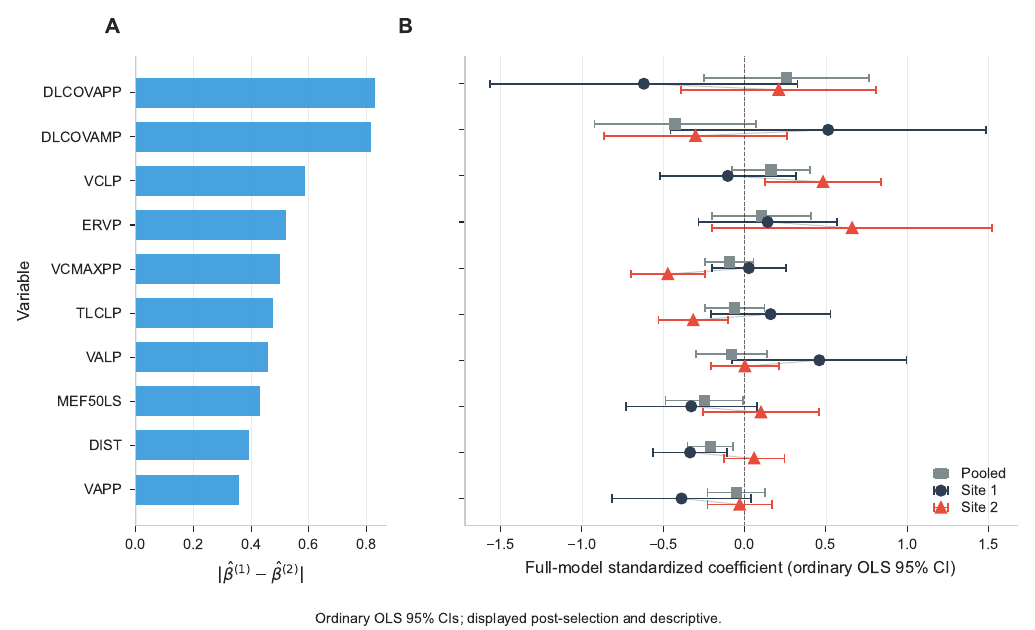}
    \caption{Original-variable coefficient heterogeneity. Panel A ranks the ten predictors with the largest absolute standardized-coefficient difference between the two site-specific 38-predictor linear models. Panel B shows the corresponding pooled and site-specific coefficients with ordinary OLS 95\% confidence intervals. Selection and estimation use the same data, so the display is descriptive and the intervals do not adjust for ranking. The analysis detects coefficient divergence but does not attribute it to case-mix or contextual sources.}
    \label{fig:original_heterogeneity}
\end{figure}

To motivate the latent coefficient-surface partition, heterogeneity was first examined entirely in the original variable space. Separate 38-predictor linear models were fitted in the two sites, and predictors were ranked by the absolute difference between their standardized coefficients. Figure~\ref{fig:original_heterogeneity} shows the ten largest differences. The ranking is independent of the latent model and of the subsequent SHAP analysis. Panel B reports ordinary OLS 95\% confidence intervals; because variables were ranked using the same coefficient estimates, the intervals are descriptive and do not provide post-selection inference. This analysis establishes that original-variable associations differ between sites, but it cannot determine whether the divergence reflects case-mix, context, or both.

\subsection{Validation of the Learned Latent Representation} \label{sec:validation}

The prognostic information retained by the six-dimensional representation was evaluated in five site-stratified observation-level folds. Within each split, scaling, PCA or autoencoder training, latent encoding, and the downstream outcome model were fitted using training rows only. Neural-network models and the random forest were evaluated with three seeds on the same five folds. As noted above, upstream imputation, correlation pruning, and outlier filtering were fixed before these splits, so the comparison is not fully leakage-free.

The standard autoencoder was trained for 500 epochs (or one epoch in the explicit under-training control) with batch size 32, learning rate $10^{-3}$, weight decay $10^{-5}$, and dropout 0.20. The proposed autoencoders used 200 epochs, batch size 180, learning rate $10^{-5}$, weight decay $10^{-6}$, and dropout 0.20. The random forest used 100 trees and maximum depth 10. These differences are reported to make clear that the standard-versus-proposed autoencoder contrast is a pipeline comparison rather than an isolated test of the loss function.

\begin{table}[h!]
\centering
\small
\caption{Held-out observation-level comparison stratified by site and conditional on the fixed upstream preprocessing. Values are means $\pm$ standard deviations; RMSE is reported on the original SGRQ-point scale. Deterministic models use five fold estimates; neural-network models and the seeded random forest use 15 fold--seed evaluations sharing the same five test folds. Standard deviations are descriptive and are not inferential uncertainty estimates. Reconstruction MSE is reported only for autoencoders.}
\label{tab:latent_validation_cv}
\begin{tabular}{p{0.40\linewidth}ccc}
\toprule
\textbf{Representation and outcome model} & \textbf{Held-out $R^2$} & \textbf{RMSE} & \textbf{Recon. MSE} \\
\midrule
Raw predictors, random forest & $0.452 \pm 0.098$ & $14.25 \pm 2.03$ & -- \\
Raw predictors, ridge regression & $0.358 \pm 0.117$ & $15.41 \pm 2.01$ & -- \\
Raw predictors, linear regression & $0.352 \pm 0.121$ & $15.47 \pm 2.04$ & -- \\
Proposed autoencoder, no native-site loss & $0.339 \pm 0.080$ & $15.63 \pm 1.49$ & $0.717 \pm 0.098$ \\
Proposed autoencoder, full objective & $0.323 \pm 0.073$ & $15.83 \pm 1.54$ & $0.694 \pm 0.051$ \\
PCA (six components), linear regression & $0.247 \pm 0.011$ & $16.74 \pm 1.68$ & -- \\
Standard autoencoder, linear regression & $0.227 \pm 0.040$ & $16.94 \pm 1.45$ & $0.543 \pm 0.030$ \\
Proposed autoencoder, no local losses & $0.216 \pm 0.049$ & $17.08 \pm 1.66$ & $0.624 \pm 0.065$ \\
Standard autoencoder, one epoch & $0.140 \pm 0.124$ & $17.84 \pm 2.03$ & $1.028 \pm 0.045$ \\
Intercept only & $-0.012 \pm 0.016$ & $19.43 \pm 2.18$ & -- \\
\bottomrule
\end{tabular}
\end{table}

The full proposed autoencoder retained more held-out prognostic information than PCA and the reconstruction-only autoencoder, but reconstructed the predictors less accurately than the standard autoencoder (Table~\ref{tab:latent_validation_cv}). Removing both local-null losses reduced mean held-out $R^2$ from $0.323$ to $0.216$ while improving reconstruction MSE from $0.694$ to $0.624$. This contrast supports the intended prognostic--reconstruction trade-off of the local objectives. Removing the native-site loss instead increased mean held-out $R^2$ to $0.339$. These predictive results do not support a claim that the native-site loss improves aggregate prognosis; its intended role is to structure site-specific coefficient surfaces, which this comparison does not test.

The fully trained standard autoencoder improved both prediction and reconstruction over its one-epoch version ($R^2$: $0.227$ vs. $0.140$; reconstruction MSE: $0.543$ vs. $1.028$). Thus, the standard-autoencoder comparator was not effectively untrained. However, the standard and proposed autoencoders also differ in optimizer settings, batch size, and training duration. Their contrast is therefore descriptive and cannot be attributed solely to the composite loss.

\subsection{Coefficient-Surface Partition and Derived Outcome-Scale Summaries} \label{sec:per_patient_signatures}

\begin{figure}[t]
    \centering
    \includegraphics[width=\linewidth]{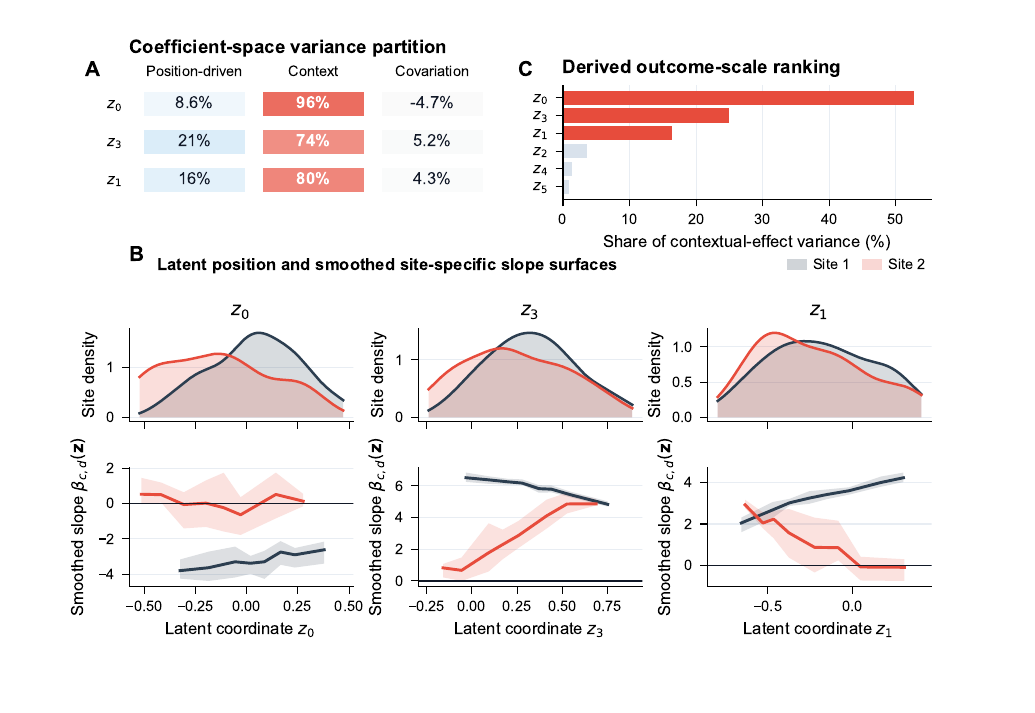}
    \caption{Coefficient-surface partition and derived outcome-scale summary for the representative seed. Panel A partitions $\mathrm{Var}_i\{\beta_r^{(c_i)}(\mathbf{z}_i)\}$ for the three coordinates with the largest total smoothed slope variance. Position-driven variation denotes changes in the reference slope across latent observation positions; context denotes site-specific departures at the same position; and covariation records whether the two patterns align or oppose each other. Position-driven variation becomes between-site case-mix heterogeneity only when the sites differ in their latent-position distributions. Panel B shows the site distributions of latent position and binned values of the smoothed own-site slope surface with within-bin interquartile ribbons. The curves are displayed against one coordinate although the surfaces are estimated in the full latent space. Panel C gives the derived outcome-scale bridge by ranking coordinates according to the variance of the slope-weighted contributions $z_{ir}\widetilde{\Delta\beta}_r$, whose coordinate-level site contrast is taken at matched latent positions as defined in the preceding text; these shares sum to 100\% across the six coordinates.}
    \label{fig:latent_dimension_sources}
\end{figure}

Figure~\ref{fig:latent_dimension_sources} presents the coefficient-surface partition for the representative encoder seed. Panel A applies the coordinate-wise variance identity from Section~\ref{sec:variance_decomp} to the three latent slopes with the largest total smoothed slope variance, separating position-driven reference-surface variation, contextual departures, and covariation. Dimensions $r=0$, $3$, and $1$ jointly account for $76.5\%$ of the summed coordinate variance $\sum_{r}\mathrm{Var}_i\{\beta_r^{(c_i)}(\mathbf{z}_i)\}$, which is the denominator for this selection statistic and is not itself the variance of any single quantity, since the coordinates need not be uncorrelated. Context accounts for $96\%$, $74\%$, and $80\%$ of the respective coordinate variances. Thus, the leading latent coefficient surfaces vary predominantly through site-specific departures from the reference surface.

Panel B shows how the sites occupy these coordinates and how their smoothed slope surfaces vary over the same positions. The sites overlap along all three coordinates, but their slope surfaces remain separated over substantial parts of that support. The curves are one-coordinate displays of surfaces estimated in the full latent space.

For the derived Panel C display only, we collapse the local coefficient-surface contrast to a single coordinate-level scalar. Both site surfaces are evaluated at every observation position and averaged over the same pooled set of positions,
\[
    \widetilde{\Delta\beta}_r
    = \frac{1}{n}\sum_{i=1}^{n}\left\{\beta_r^{(1)}(\mathbf{z}_i)-\beta_r^{(2)}(\mathbf{z}_i)\right\}.
\]
Comparing the two surfaces at the same positions cancels the reference surface pointwise, so only site-specific departures survive; at $C=2$ with $q_k=1/2$ this scalar equals $2n^{-1}\sum_i\beta_r^{(1),\mathrm{ctx}}(\mathbf{z}_i)$, built from the same surfaces as the primary partition. Averaging each site over its own positions instead would retain the difference in reference-surface means between the site position distributions, letting a purely case-mix difference appear as a site-specific one. Panel C provides the derived outcome-scale bridge across dimensions. It ranks the variance of the slope-weighted contributions $z_{ir}\widetilde{\Delta\beta}_r$ used in the descriptive contextual-effect summary. Dimensions $r=0$, $3$, and $1$ contribute $52.7\%$, $24.9\%$, and $16.4\%$, together accounting for $94.0\%$ of the summed coordinate variance $\sum_{r}\mathrm{Var}_i\{z_{ir}\widetilde{\Delta\beta}_r\}$. The same dimensions therefore dominate the derived outcome-scale summary and show predominantly contextual variation in the primary coefficient-space partition. The scalar approximates the position-varying departure $\beta_r^{(c_i),\mathrm{ctx}}(\mathbf{z}_i)$ by one value per coordinate. Recomputing with that departure leaves the ordering unchanged but moves the shares to $37.1\%$, $24.4\%$, and $20.8\%$, summing to $82.3\%$. The ordering is therefore robust to the approximation while the magnitudes are not, and the scalar concentrates more variance in the leading coordinate than the position-varying departure does.

Projecting the coefficient-surface partition onto the outcome scale gives two scalars for each eligible observation. The case-mix component, $M_i^{\mathrm{mix}}$, captures the model-implied outcome component at that observation's latent position under the cross-site reference relationship. The contextual component, $M_i^{\mathrm{ctx}}$, captures the site-specific shift at the same latent position.

The case-mix component overlaps substantially between sites. The median $M_i^{\mathrm{mix}}$ is $-0.25$ in Site~1 (IQR $[-0.56, +0.06]$) and $-0.31$ in Site~2 (IQR $[-0.62, +0.26]$). Thus, observation-level latent position implies similar model-implied outcome ranges across both sites.

The contextual component shows a clearer site separation. The median $M_i^{\mathrm{ctx}}$ is $+0.17$ in Site~1 (IQR $[0.00, +0.30]$) and $-0.23$ in Site~2 (IQR $[-0.44, -0.12]$). At a matched latent position and $C = 2$, the contextual intercept and slope departures have opposite signs across sites by construction because the reference surfaces $\bar{\alpha}$ and $\bar{\boldsymbol{\beta}}$ assign equal weight to each site. The observed median separation shows that this antisymmetry aligns with the empirical observation distributions. The magnitude $|M_i^{\mathrm{ctx}}|$, rather than the sign, is the informative observation-level quantity.

Magnitude-based stratification gives a site-balanced view of the strongest observation-level effects. Among the top decile by $|M_i^{\mathrm{ctx}}|$ (30 observations), 15 observations originate from Site~1 and 15 from Site~2. Among the top decile by $|M_i^{\mathrm{mix}}|$ (30 observations), 13 are from Site~1 and 17 from Site~2. This descriptive stratification identifies observations for which each source most strongly influences the model-implied outcome component.

Together, these components define the derived observation-level diagnostic: case-mix is the position-driven part of the outcome surface, while context is the site-specific shift at that same position. Aggregated results from 100 observation-level bootstrap-OOB replicates of the representative encoder seed show that the variance of $M_i$ is case-mix-leading. $\mathrm{Var}_i(M_i^{\mathrm{mix}})$ contributes $63\% \pm 8\%$, $\mathrm{Var}_i(M_i^{\mathrm{ctx}})$ contributes $41\% \pm 9\%$, and $2\,\mathrm{Cov}_i(M_i^{\mathrm{mix}}, M_i^{\mathrm{ctx}})$ contributes $-4\% \pm 9\%$, summing to $100\%$ by design.

The case-mix-leading outcome-scale pattern is stable across encoder initializations, although not perfectly so. Across 10 trained encoders, per-seed mean case-mix shares range from $60\%$ to $84\%$, and context shares from $19\%$ to $44\%$. All 10 seeds have a larger case-mix than context share at the per-seed point estimate, henceforth termed case-mix-leading. At the bootstrap-replicate level, a small fraction, up to approximately 10\% in the most variable seeds, crosses the case-mix/context boundary. Thus, residual encoder-level instability is reduced but not eliminated.

In this descriptive ten-seed comparison, gradient-scaled fits varied less across encoder initializations (Section~\ref{sec:composite_loss}). Without this adjustment, seven of ten seeds were case-mix-leading and three were context-leading, with per-seed case-mix shares ranging from $38\%$ to $87\%$ and context shares from $22\%$ to $69\%$. With gradient scaling, these range widths narrowed by about one half on both axes while preserving the qualitative result that both sources contribute substantially to outcome-scale variation. This comparison describes the observed stability difference but does not establish its mechanism.

\subsubsection{Between-site aggregation of the outcome-scale partition}

We now change the unit of summary from variation across observations to variation between site means. The within/between aggregation in Section~\ref{sec:outcome_decomp} makes this distinction explicit. Applied to $\mathrm{Var}_i(M_i)$, the between-site component is $\tau_M^2$; at $C = 2$, it reduces to $\pi_1\pi_2(\bar{M}_1 - \bar{M}_2)^2 = 14.6\%$ of $\mathrm{Var}_i(M_i)$.

Table~\ref{tab:within_between} reports the full split for the representative seed. The case-mix component is almost entirely within-site ($63.3\%$ within vs. $0.24\%$ between), consistent with the similar site-specific case-mix medians reported above. The contextual component divides nearly evenly ($20.1\%$ within vs. $18.6\%$ between). In the representative seed, this near-equality shows that the contextual contribution is not exhausted by the average site shift: it also varies across observations within a site. This within-site variation does not by itself establish that the departure surfaces vary with latent position, because $M_i^{\mathrm{ctx}}$ varies with $\mathbf{z}_i$ even when $\boldsymbol{\beta}^{(c),\mathrm{ctx}}$ is constant. The seed-averaged path shows the same near-$50/50$ split ($9.7\%$ within vs. $10.2\%$ between of seed-averaged $\mathrm{Var}_i(M_i)$), so the pattern is not specific to the representative seed.

\begin{table}[h!]
\centering
\caption{Within/between aggregation of the outcome-scale variance partition (representative seed point estimate; the preceding paragraph's $41\% \pm 9\%$ context share is the bootstrap mean over replicates of the same seed), as percentages of $\mathrm{Var}_i(M_i)$. For $M_i^{\mathrm{ctx}}$, the between-site row corresponds to the average site shift, while the within-site row corresponds to observation-level variation in that shift. The negative between-site covariation indicates that the site with higher mean case-mix has lower mean context.}
\label{tab:within_between}
\begin{tabular}{lrr}
\toprule
\textbf{Component} & \textbf{Within-site} & \textbf{Between-site} \\
\midrule
Case-mix component   & $63.3\%$ & $0.24\%$ \\
Context component    & $20.1\%$ & $18.6\%$ \\
$2\,\mathrm{Cov}_i(M_i^{\mathrm{mix}}, M_i^{\mathrm{ctx}})$ & $+1.9\%$ & $-4.2\%$ \\
\midrule
Total                & $85.3\%$ & $14.6\%$ \\
\bottomrule
\end{tabular}
\end{table}

The permuted-site negative control tests whether the pipeline produces a context partition when site labels carry no signal. The training and bootstrap-OOB protocol was repeated after uniformly permuting site labels (Appendix, \emph{Permuted-site negative control}). Under this null, the between-site case-mix share is $0.79\% \pm 0.43\%$ and the between-site context share is $1.37\% \pm 0.83\%$ of $\mathrm{Var}_i(M_i)$. The real-data between-site case-mix share is $0.24\%$, whereas the real-data context share is $18.6\%$. The latter substantially exceeds the permuted-label summary. The across-observation context share also drops by a factor of about three under permutation, with the covariation share collapsing to zero. Together, these descriptive contrasts are consistent with a site-level contextual signal rather than an artifact recovered from arbitrary labels.

A separate control asks whether the asymmetry between case-mix and context depends on training the encoder on site labels at all. This check applies the same outcome-scale partition and within/between aggregation to the latent space of a reconstruction-only autoencoder, which by construction does not see site labels during training. Across 10 seeds (single-fit on the full sample, otherwise matching the production $\sigma$ and $k_{\mathrm{NN}}$), the asymmetry persists: between-site case-mix is $0.59\% \pm 0.75\%$ of $\mathrm{Var}_i(M_i)$, sitting within the permuted-label null band, whereas between-site context is $9.5\% \pm 2.8\%$, well above the corresponding null. The asymmetry therefore appears to reflect structure in the data rather than a property of site-aware training. The production training has a larger between-site context share than the reconstruction-only baseline ($18.6\%$ versus $9.5\%$), but this descriptive comparison does not establish the mechanism underlying that difference.

\subsection{Exploratory Prediction-error Diagnostic} \label{sec:impact}

The coefficient-surface partition attributes heterogeneity to position-driven and site-specific sources. Its outcome-scale projection summarizes their contributions. As a complementary exploratory check, the analysis examined the loss of accuracy when observations from each site were predicted using the other site's model at matched latent locations. If contextual model differences are present, swapping site-specific models should produce a larger MSE than the own-site assignment. This diagnostic assesses a different summary from the coefficient and outcome-scale partitions, but it is not an independent validation.

The check used 100 site-stratified observation-level bootstrap replicates for each of 10 encoder seeds. In each replicate, the bootstrap sample defined an out-of-bag set. Within that set, a Ridge regression with penalty 0.1 was fitted for each site to map the latent representation $\mathbf{Z}$ to the outcome. Under the own-site assignment, observations were predicted by their site's model; under the swapped assignment, they were predicted by the other site's model.

Across the 1,000 seed--replicate estimates, swapping increased observation-level MSE by a mean of $99.7\%$ (SD $40.4\%$). This large descriptive penalty is consistent with site-conditional differences in the $\mathbf{z}{\to}y$ relationship, but it should not be interpreted as independent validation of the coefficient-surface partition.

\subsection{Interpretation of the Contextual Effect} \label{sec:context_traceback}

Finally, contextual coefficient-surface patterns were linked descriptively to the original clinical predictors. SHAP (SHapley Additive exPlanations) analysis was applied directly to the encoder network \citep{lundbergUnifiedApproachInterpreting2017}. This yields per-feature, per-coordinate attributions that quantify how each input predictor contributes to each latent coordinate $z_r$. These encoder attributions do not transform the latent slopes into original-variable regression coefficients.

For the descriptive SHAP summary, each feature's attribution to a latent coordinate was weighted by the matched-position scalar contrast $\widetilde{\Delta\beta}_r$ defined above. The resulting slope-weighted attribution is a descriptive outcome-scale summary connecting input predictors to contextual latent-slope patterns. The primary coefficient-surface partition remains defined on the full smoothed surfaces $\alpha^{(k)}(\mathbf{z})$ and $\boldsymbol{\beta}^{(k)}(\mathbf{z})$ and on their contextual departures $\alpha^{(k),\mathrm{ctx}}(\mathbf{z})$ and $\boldsymbol{\beta}^{(k),\mathrm{ctx}}(\mathbf{z})$, which the scalar summarizes by one value per coordinate.

\begin{figure}
    \centering
    \includegraphics[width=\linewidth]{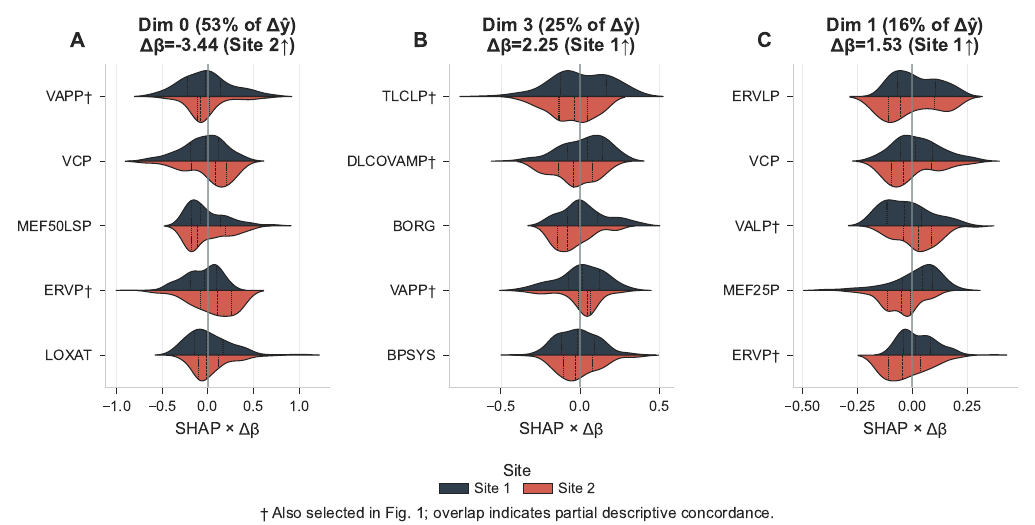}
    \caption{Original-predictor trace-back of the contextual latent-slope pattern. Each panel shows the site-specific distribution of encoder SHAP values weighted by the coordinate-level matched-position slope contrast $\widetilde{\Delta\beta}_r$ for the five largest input-predictor attributions. The three coordinates contribute 52.7\%, 24.9\%, and 16.4\% of the slope-weighted variance. These weighted encoder attributions are descriptive links from input predictors to latent coefficient patterns, not original-variable coefficient estimates. Daggers mark predictors that also occur in the independent Figure~\ref{fig:original_heterogeneity} top-ten ranking; this overlap indicates partial descriptive concordance, not independent validation. All latent artifacts, encoder weights, and the deterministic SHAP background correspond to representative seed 1213.}
    \label{fig:encoder_shap}
\end{figure}

Figure~\ref{fig:encoder_shap} traces the three leading contextual dimensions back to the original predictors using the representative seed 1213 for the latent artifacts, encoder, and SHAP background. Coordinate 0 accounts for 52.7\% of the slope-weighted variance and has a negative matched-position slope contrast ($\widetilde{\Delta\beta}_0=-3.44$); its five largest weighted attributions are VAPP, VCP, MEF50LSP, ERVP, and LOXAT. Coordinate 3 contributes 24.9\% with $\widetilde{\Delta\beta}_3=+2.25$ and is associated most strongly with TLCLP, DLCOVAMP, BORG, VAPP, and BPSYS. Coordinate 1 contributes 16.4\% with $\widetilde{\Delta\beta}_1=+1.53$ and is associated with ERVLP, VCP, VALP, MEF25P, and ERVP.

Daggers in Figure~\ref{fig:encoder_shap} mark variables that also appear among the independent original-space coefficient differences in Figure~\ref{fig:original_heterogeneity}. Seven of the 15 displayed dimension--feature instances overlap, representing five distinct variables: VAPP, ERVP, TLCLP, DLCOVAMP, and VALP. This partial overlap links the latent coefficient-surface patterns to heterogeneity visible in the original predictors. It does not establish a correspondence between latent and original-variable coefficients and is not independent validation, because both analyses use the same observations.

\section{Discussion} \label{sec:discussion}

In synthesizing prognostic models across sites, researchers must decide whether to pool estimates globally or to synthesize site-specific models. Standard regression on the original variables flags the existence of coefficient heterogeneity (Figure~\ref{fig:original_heterogeneity}), but it does not identify the channel through which the heterogeneity arises. The proposed approach addresses this gap through a partition of learned local coefficient surfaces into position-driven reference variation and site-specific departures. This primary partition is structurally closest to the coefficient heterogeneity familiar from meta-analysis, although its coefficients refer to learned latent coordinates. Projection then gives an outcome-scale case-mix/context partition, and its between-site aggregation, $\tau_M^2$, provides a descriptive analogue to a meta-analytic heterogeneity measure rather than a replacement for a coefficient-level $\tau^2$ estimator. This distinction matters for synthesis: case-mix heterogeneity can often be addressed through standardization, whereas contextual heterogeneity points toward site-specific or context-aware synthesis.

The decision consequence depends on whether synthesis is symmetric or asymmetric. In a symmetric analysis, all sites are placed on equal footing: case-mix differences motivate standardization to a common population or a sufficiently flexible pooled surface, whereas persistent contextual differences motivate retaining site-specific relationships through stratification, partial pooling, or random-effects synthesis. In an asymmetric analysis, one site is the target: case-mix differences can in principle be addressed by transporting or reweighting source information to that target, while contextual differences indicate that the source relationship itself may not transport without target-specific recalibration or outcome data. The framework diagnoses which concern is present; it does not by itself select a final estimator or borrowing rule.

In the COPD application, context accounts for $74\%$--$96\%$ of the coefficient-surface variation in the three leading latent slopes. The derived summaries then change the reading. Across observations, the outcome-scale variance partition is case-mix-leading: latent position under the reference surface accounts for the larger share of variation. At the between-site aggregation, however, case-mix is near the permuted-label null floor, while context is well above it. Thus, the primary coefficient-space result and the site contrast relevant for synthesis both indicate site-conditional differences, even though the across-observation outcome-scale summary is case-mix-leading (Section~\ref{sec:per_patient_signatures}). The exploratory observation-level model-swap check provides a complementary prediction-error signal (Section~\ref{sec:impact}).

At the observation level, the covariation component is small ($-4\% \pm 9\%$) under gradient-scaled training (Section~\ref{sec:composite_loss}). Thus, the covariance between $M_i^{\mathrm{mix}}$ and $M_i^{\mathrm{ctx}}$ contributes little to total variance in the fitted outcome-scale partition. In settings with larger covariation, reweighting case-mix alone would be less interpretable because the two channels would act through the same latent regions. Here, the main synthesis concern is instead a between-site contextual shift that persists after the near-null between-site case-mix term is accounted for.

The methodological contribution integrates local regression with representation learning to support a coefficient-surface partition and a derived outcome-scale variance summary. Under measured common support and an adequate learned representation, site-specific local models distinguish position-driven variation from differences between sites at matched profiles.

Local regression captures non-linearity and effect modification without requiring pre-specified functional forms. The reconstruction objective makes these local models feasible in a high-dimensional predictor space by reducing dimensionality. Joint training with the local-fit loss then shapes the latent geometry around prognostic relationships rather than reconstruction alone.

The coefficient-surface partition supplies the source attribution, while the outcome-scale projection and $\tau_M^2$ summarize its observation- and site-level implications. Together, these outputs inform the evaluation of pooled versus site-specific synthesis. When case-mix dominates and contextual heterogeneity is minimal, pooled synthesis after standardization may be adequate. When site-conditional coefficient differences are substantial, as observed here, context-aware or site-specific synthesis should be evaluated even if the derived across-observation partition is case-mix-leading.

This work complements existing approaches to case-mix heterogeneity. Methods such as model-based standardization and causally interpretable meta-analysis \citep{wangCausalMetaR2025} enable the transport of estimates to a target population when relevant moderators are known and measured. The proposed framework addresses a different scenario: diagnosing the source of unexplained heterogeneity when moderators are unknown or when the predictor space is too high-dimensional for explicit interaction modeling. Rather than replacing standardization approaches, this method assists researchers in determining whether standardization is likely to be effective or whether contextual factors require further investigation before valid synthesis can be achieved.

Several limitations concern inferential scope and tuning. The coefficient-space partition is conditional on the scale and orientation of the learned latent representation. Its slopes are not original-variable coefficients, and rotated or reflected axes cannot be compared dimension by dimension across independently trained encoders without alignment. The outcome-scale projection provides a scalar summary that is less tied to individual coordinates, but neither level formally tests whether one component is larger than another. Both depend on the latent dimension size ($D$), loss weights ($\lambda_{\mathrm{rec}}$, $\lambda_{\mathrm{local}}$, and $\lambda_{\mathrm{native}}$), neighborhood fraction ($k_{\mathrm{NN}}$), kernel scale ($\sigma$), and gradient scaling $\eta$. The Appendix reports the local-regression bias--variance selection trade-off, but sensitivity across diverse datasets remains to be assessed.

The empirical setting imposes further limits. The analysis used two sites, the minimum required for a between-site variance contrast, and therefore cannot show how contextual patterns vary across multiple sites. The findings also come from one clinical dataset; external validation is needed in settings where one source clearly dominates on both the variance and prediction-error axes. In addition, all analyses are conditional on fixed upstream preprocessing. K-nearest-neighbor imputation, including the contemporaneous outcome, correlation pruning, and outlier filtering were performed once on the full dataset. The observation-level held-out comparison assigns observations directly to folds, so this upstream processing means it is not fully leakage-free.

The validation summaries should be interpreted descriptively. Fold--seed standard deviations share five test folds and are not inferential. The observation-level model-swap bootstrap is exploratory and does not provide independent validation. Figure~\ref{fig:encoder_shap} also uses the same observations as Figure~\ref{fig:original_heterogeneity}; their overlap is a trace-back within this dataset, not independent confirmation.

Finally, IPD are often difficult to obtain in practice. Systematic reviews report that only a minority of IPD meta-analyses retrieve data from all eligible studies \citep{nevittExploringChangesTime2017, polaninEffortsRetrieveIPD2018}. The framework is a diagnostic on the available IPD subset, but its practical value depends on assembling enough sites with IPD to support the analysis.

The framework could be extended to outcomes commonly encountered in IPD meta-analysis, including longitudinal and time-to-event outcomes. It also generates a "contextual signature" for each site: the pair $\{\alpha^{(k),\mathrm{ctx}}(\mathbf{z}),\boldsymbol{\beta}^{(k),\mathrm{ctx}}(\mathbf{z})\}$ that characterizes how prognostic relationships differ from the reference surfaces across latent profiles. Future research could summarize these signatures in a second-stage meta-regression to identify site-level characteristics, such as hospital volume or measurement protocols, that are associated with contextual heterogeneity.

The representation itself also warrants further evaluation. Conditional on the fixed upstream preprocessing, it retained more held-out prognostic information than PCA or a reconstruction-only autoencoder while remaining in the range of raw-feature linear models. This supports the outcome-guided objective, but leaves open whether a linear encoder with the same loss could achieve comparable results.

Observation-level case-mix and context signatures might eventually support subgroup-targeted analyses or stratified borrowing. Any such rule would require independent external evaluation. That evaluation, and scalability to meta-analyses involving many sites, is beyond the scope of this paper.

In conclusion, this approach partitions learned coefficient surfaces into a shared position-driven component, site deviations, and their covariation. Projection onto the outcome scale and subsequent within/between aggregation provide derived scalar summaries. In the COPD application, contextual differences dominate the leading latent coefficient surfaces. The derived across-observation outcome-scale partition is case-mix-leading, whereas its between-site aggregation is context-dominated and the context share substantially exceeds the permuted-label summary. An exploratory observation-level model swap provides a concordant prediction-error signal. The method does not select a final synthesis strategy, but it makes its model-based source attribution explicit when pooled and site-specific models are being considered. When only a subset of studies provide IPD, the framework can flag potential concerns for the broader meta-analysis, conditional on the learned representation and on the sites with available IPD representing the remaining evidence. Extrapolation to aggregate-data studies requires additional assumptions and external evidence.

\vspace*{1pc}

\noindent {\bf{Author Contributions}}

\noindent \textbf{Max Behrens:} Conceptualization, Formal analysis, Methodology, Software, Validation, Visualization, Writing -- original draft. \textbf{Janis M. Nolde:} Conceptualization, Writing -- review \& editing. \textbf{Eleni Papakonstantinou:} Data curation, Funding acquisition, Resources, Writing -- review \& editing. \textbf{Gabriele Bellerino:} Methodology, Writing -- review \& editing. \textbf{Theodoros Evrenoglou:} Methodology, Writing -- review \& editing. \textbf{Angelika Rohde:} Methodology, Funding acquisition, Writing -- review \& editing. \textbf{Daiana Stolz:} Data curation, Funding acquisition, Resources, Writing -- review \& editing. \textbf{Moritz Hess:} Conceptualization, Data curation, Methodology, Software, Supervision, Writing -- review \& editing. \textbf{Harald Binder:} Conceptualization, Funding acquisition, Methodology, Supervision, Writing -- review \& editing.

\vspace*{1pc}

\noindent {\bf{Ethics Approval and Consent}} This work is a secondary analysis of the PREVENT trial \citep{stolzIntensifiedTherapyInhaled2018}, registered as ISRCTN45572998. The trial was approved by the institutional review boards of the participating centers under reference EKBB 306/10, and all patients gave written informed consent. It was conducted in accordance with the Declaration of Helsinki and the guidelines on good clinical practice. No additional patient contact or data collection was undertaken for the present analysis.

\vspace*{1pc}

\noindent {\bf{Data Availability}} The patient-level PREVENT data analyzed here are not publicly available. They were collected under the trial's consent and governance arrangements, which do not permit redistribution, and access requests are handled by the PREVENT trial investigators \citep{stolzIntensifiedTherapyInhaled2018}. The aggregate results reported in this article, including the variance partitions and the between-site aggregation, are derived from those data and are reported in the article and its appendix. The figures additionally display per-observation quantities, which are not tabulated.

\vspace*{1pc}

\noindent {\bf{Code Availability}} All analysis code is available at \url{https://github.com/maxjonasbehrens/case-mix-context-decomposition} under the MIT license. The repository ships a synthetic example dataset generated from a fixed seed, so the full decomposition pipeline can be run end to end, and the aggregate outputs reproduced, without access to the patient-level data.

\vspace*{1pc}

\noindent {\bf{Acknowledgement}} This work was funded by the Deutsche Forschungsgemeinschaft (DFG, German Research Foundation) – Project-ID 499552394 – SFB 1597 (MB, JMN, GB, AR, MH and HB). JMN is supported by the Berta-Ottenstein-Programme for Clinician Scientists, Faculty of Medicine, University of Freiburg. TE was supported by the Deutsche Forschungsgemeinschaft (DFG, German Research Foundation) under Project ID 554095932.
\vspace*{1pc}

\noindent {\bf{Conflict of Interest}}

\noindent {\it{The authors have declared no conflict of interest.}}

\section*{Highlights}

\subsection*{What is already known?}
\begin{itemize}
    \item Between-site heterogeneity in prognostic associations may reflect differences in case mix, conditional outcome relationships, or both, with different implications for evidence synthesis.
    \item Conventional coefficient-level heterogeneity measures summarize disagreement but do not decompose it into variation associated with patient profiles, site-specific context, and their covariation.
\end{itemize}

\subsection*{What is new?}
\begin{itemize}
    \item We introduce a framework that fits site-specific local regressions in a dimension-reduced space and partitions the resulting prognostic relationships into a position-driven reference component and a contextual component that compares sites at comparable patient profiles, together with their covariation in the corresponding variance identity.
    \item Derived observation- and site-level summaries translate this decomposition to the outcome scale, connecting the primary coefficient-scale attribution to heterogeneity across observations and between sites.
\end{itemize}

\subsection*{Potential impact for RSM readers}
\begin{itemize}
    \item The diagnostic identifies whether population alignment and standardization or site-specific relationships and model adaptation warrant further assessment.
    \item At sites contributing IPD, it can reveal contextual departures relevant to syntheses that combine IPD and aggregate data, conditional on overlapping patient profiles (common support) and on how well those sites represent settings without IPD.
\end{itemize}

\bibliographystyle{elsarticle-harv} 
\bibliography{main}

\section*{Appendix}

\subsection*{Autoencoder Architecture}

We utilized an autoencoder neural network to perform non-linear dimensionality reduction. The encoder function maps the high-dimensional input vector (dimension $p=38$) to a lower-dimensional latent representation ($D=6$). This mapping is achieved through a series of transformations that progressively compress the data structure: the input is passed through intermediate layers of size 24 and 18 before reaching the bottleneck layer. We employed hyperbolic tangent (Tanh) non-linear activation functions to capture complex dependencies, with dropout regularization (rate 0.20). A symmetric decoder attempts to reconstruct the original covariates from this latent representation, ensuring the latent space retains the relevant information.

\subsection*{Hyperparameter Configuration}

Table~\ref{tab:hyperparameters} summarizes the hyperparameter settings used in the experiments.

\begin{table}[h!]
\centering
\caption{Hyperparameter configuration for the production model.}
\label{tab:hyperparameters}
\begin{tabular}{llc}
\toprule
\textbf{Category} & \textbf{Parameter} & \textbf{Value} \\
\midrule
\multirow{4}{*}{Architecture} & Input dimension ($p$) & 38 \\
& Latent dimension ($D$) & 6 \\
& Hidden layers & 24 $\rightarrow$ 18 $\rightarrow$ 6 \\
& Dropout rate & 0.20 \\
\midrule
\multirow{4}{*}{Loss weights} & $\lambda_{\text{rec}}$ (reconstruction) & 1.0 \\
& $\lambda_{\text{local}}$ (local fit) & 0.5 \\
& $\lambda_{\text{native}}$ (native-site loss) & 0.5 \\
& $\eta$ (first-layer gradient scaling on $\text{Loss}_{\text{native}}$) & 0.1 \\
\midrule
\multirow{3}{*}{Local regression} & Neighborhood fraction ($k_{\mathrm{NN}}$) & 0.25 \\
& Kernel bandwidth ($\sigma$) & 0.3 \\
& WLS ridge stabilizer ($\lambda_{\mathrm{ridge}}$) & $1 \times 10^{-4}$ \\
\midrule
\multirow{2}{*}{Coefficient-surface smoother} & KernelRidge RBF parameter ($\gamma$) & 0.1 \\
& KernelRidge ridge penalty ($\lambda_{\mathrm{KRR}}$) & 1.0 \\
\midrule
\multirow{4}{*}{Training} & Learning rate & $1 \times 10^{-5}$ \\
& Epochs & 200 \\
& Batch size & 180 \\
& LR reduction patience & 20 epochs \\
\bottomrule
\end{tabular}
\end{table}

\subsection*{Localized Regression Parameters}

The local regression framework requires two key hyperparameters: the neighborhood fraction $k_{\mathrm{NN}}$ (which determines the site-specific neighbor count $m_k$ for each local fit) and the kernel bandwidth $\sigma$ (which controls how rapidly influence decays with distance in the latent space). These parameters jointly determine the effective smoothing of the local coefficient estimates and involve a bias-variance tradeoff.

A small neighborhood with narrow bandwidth yields highly localized estimates that can capture fine-grained variation in prognostic relationships, but may be unstable due to limited effective sample size. Conversely, a large neighborhood with wide bandwidth produces stable estimates that may obscure genuine local variation. To select appropriate values, we evaluated six neighborhood fractions (0.10 to 0.35 in steps of 0.05) and six kernel scales (0.2 to 0.7 in steps of 0.1), giving 36 configurations, using a vanilla autoencoder (trained only on reconstruction loss) to isolate the effect of these parameters from the representation learning.

\begin{figure}
    \centering
    \includegraphics[width=0.9\linewidth]{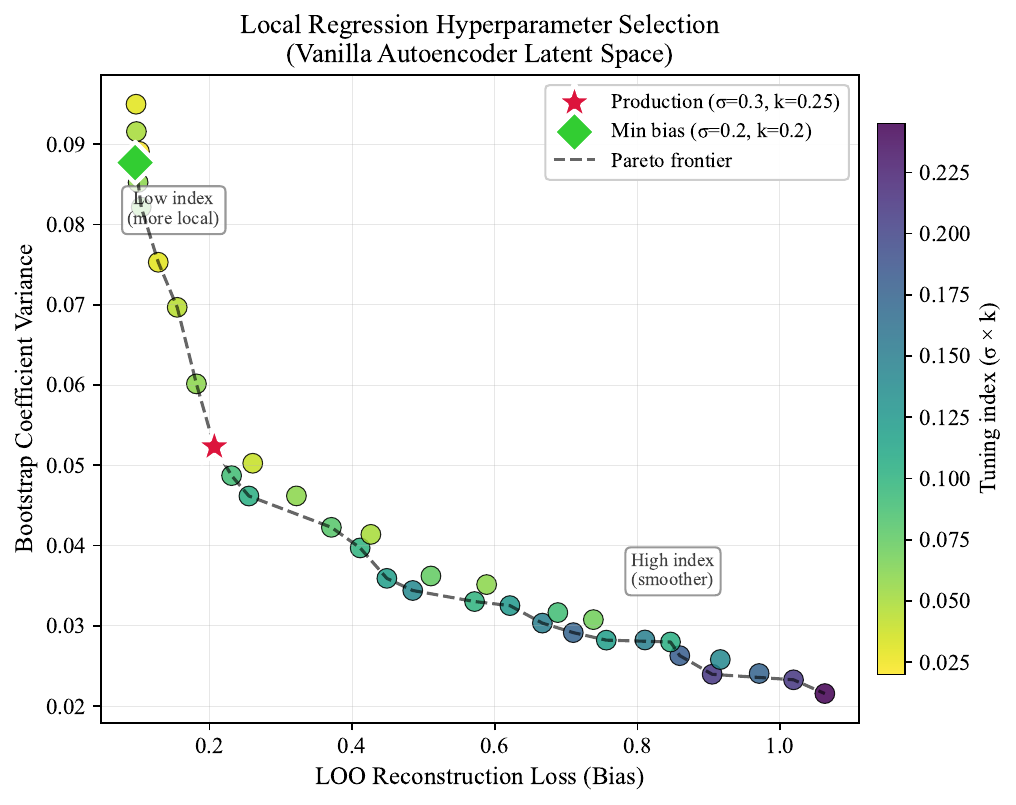}
    \caption{Bias--variance trade-off for local regression hyperparameters evaluated in a vanilla-autoencoder latent space trained only on reconstruction loss. The x-axis shows leave-one-out latent reconstruction error, the squared distance between a held-out observation's latent position and the kernel-weighted mean of its neighbors' positions, which measures neighborhood smoothing bias without involving the outcome. The y-axis shows bootstrap coefficient variance under neighbor resampling. Color represents the tuning index $\sigma \times k_{\mathrm{NN}}$; the effective distance bandwidth for a given query is $\sigma b_i^{(k)}$. Production fits use the selected fraction with site-specific neighbor counts. Light points yield lower reconstruction error but greater variance, while dark points are more stable but have greater reconstruction error. The selected production configuration ($\sigma=0.3$, $k_{\mathrm{NN}}=0.25$; red star) balances these criteria. The dashed Pareto frontier comprises configurations for which neither metric can improve without worsening the other.}
    \label{fig:local_tradeoff}
\end{figure}

For each configuration, we computed two metrics. Leave-one-out latent reconstruction error measured neighborhood smoothing bias as the squared distance between a held-out observation's latent position and the kernel-weighted mean of its neighbors' positions; it is a property of the latent geometry and does not involve the outcome. Bootstrap coefficient variance measured stability under neighbor resampling. Figure~\ref{fig:local_tradeoff} displays this trade-off. We selected $k_{\mathrm{NN}}=0.25$ and $\sigma=0.3$ from the Pareto frontier to prioritize low reconstruction error while maintaining acceptable coefficient variance. This choice reflects our preference for detecting local variation rather than smoothing it away. In the production implementation, local fits are restricted to observations from the same site, and the bandwidth distance is selected at zero-based index $\lfloor 0.25 n_{k,\mathrm{ref}}\rfloor$ among the sorted reference-set distances for site $k$.

\subsection*{Model Training}

Parameters were estimated via stochastic optimization using the Adam algorithm \citep{kingmaAdamMethodStochastic2017}. The optimization process ran for 200 complete passes through the dataset (epochs), updating estimates using mini-batches of 180 observations. The global objective function minimized a weighted sum of three components corresponding to the composite loss defined in the Methods: the reconstruction loss $\text{Loss}_{\text{rec}}$ (weight $\lambda_{\text{rec}}=1.0$), the local fit loss $\text{Loss}_{\text{local}}$ (weight $\lambda_{\text{local}}=0.5$), and the native-site loss $\text{Loss}_{\text{native}}$ (weight $\lambda_{\text{native}}=0.5$). The gradient flowing from $\text{Loss}_{\text{native}}$ into the first encoder block was scaled by $\eta=0.1$ via straight-through gradient scaling; the other two losses updated the first layer at full gradient. Learning rate adaptation was handled automatically: if the average training loss did not improve for 20 epochs, the learning rate (initially $1 \times 10^{-5}$) was reduced by a factor of 10.

\subsection*{Permuted-site negative control}

To assess whether the derived outcome-scale variance partition could result from a pipeline applied to labels lacking genuine site signal, site labels were permuted uniformly at random across the full dataset (preserving per-site sample sizes), and the training and bootstrap-OOB protocol was repeated. Across 10 seeds with 100 replicates each, the across-observation context share decreased to $14.1\% \pm 2.8\%$ (range $9.1\%$--$17.9\%$), the covariation share collapsed to $0.2\% \pm 7.8\%$ centered on zero, and all 10 seeds remained case-mix-leading at the per-seed point estimate, compared to $41\% \pm 9\%$ and $-4\% \pm 9\%$ under the real labels. Case-mix increased to $85.7\% \pm 9.4\%$ due to a structural effect: under the null, the across-observation outcome variance decreases more rapidly than the case-mix component, causing its proportion to increase by construction. Taking Site~1 as the target, the transport contrast for observation $i$ is $\Delta_i^{\mathrm{target}}=\alpha^{(1)}(\mathbf{z}_i)-\alpha^{(c_i)}(\mathbf{z}_i)+\mathbf{z}_i^T\{\boldsymbol{\beta}^{(1)}(\mathbf{z}_i)-\boldsymbol{\beta}^{(c_i)}(\mathbf{z}_i)\}$. Its SD across Site~2 observations decreased from approximately $0.7$ under real labels to a seed-averaged $0.51 \pm 0.08$ under the null.

\subsection*{Predictor Variables}

Table~\ref{tab:variables} lists the 38 predictor variables used in the analysis. Several physiological quantities appear in multiple representations (absolute units, percent predicted, litres predicted); each was retained as a separate input. Canonical definitions follow the PREVENT study protocol \citep{stolzIntensifiedTherapyInhaled2018}.

\begin{longtable}{p{0.22\linewidth}p{0.68\linewidth}}
\caption{Predictor variables from the PREVENT study ($p = 38$).}
\label{tab:variables}\\
\toprule
\textbf{Variable} & \textbf{Description} \\
\midrule
\endfirsthead
\multicolumn{2}{l}{\tablename\ \thetable\ (continued)}\\
\toprule
\textbf{Variable} & \textbf{Description} \\
\midrule
\endhead
\multicolumn{2}{l}{\textit{Spirometry \& expiratory flow}} \\
VCMAXL    & Maximum vital capacity (L) \\
VCMAXPP   & Maximum vital capacity (\% predicted) \\
VCP       & Vital capacity (\% predicted) \\
VCLP      & Vital capacity (L predicted) \\
FVCLP     & Forced vital capacity (L predicted) \\
PEFLS     & Peak expiratory flow (L/s) \\
MEF50LS   & Mid-expiratory flow at 50\% FVC (L/s) \\
MEF50P    & Mid-expiratory flow at 50\% FVC (\% predicted) \\
MEF50LSP  & Mid-expiratory flow at 50\% FVC (L/s predicted) \\
MEF25P    & Mid-expiratory flow at 25\% FVC (\% predicted) \\
\midrule
\multicolumn{2}{l}{\textit{Lung volumes}} \\
TLCP      & Total lung capacity (\% predicted) \\
TLCLP     & Total lung capacity (L predicted) \\
RVL       & Residual volume (L) \\
RVTLCP    & Residual volume / total lung capacity ratio (\%) \\
ERVL      & Expiratory reserve volume (L) \\
ERVP      & Expiratory reserve volume (\% predicted) \\
ERVLP     & Expiratory reserve volume (L predicted) \\
VALP      & Alveolar volume (L predicted) \\
VAPP      & Alveolar volume (\% predicted) \\
\midrule
\multicolumn{2}{l}{\textit{Gas transfer}} \\
DLCOVAPP  & Diffusing capacity for carbon monoxide (\% predicted) \\
DLCOVAMP  & Diffusing capacity per alveolar volume (\% predicted) \\
NO1       & Exhaled nitric oxide \\
\midrule
\multicolumn{2}{l}{\textit{Cardiovascular \& vital signs}} \\
BPSYS     & Systolic blood pressure \\
BPDIA     & Diastolic blood pressure \\
HR        & Heart rate (resting) \\
HHR       & Heart rate (alternate measurement) \\
BREATH    & Respiratory rate \\
\midrule
\multicolumn{2}{l}{\textit{Oxygenation \& exercise capacity}} \\
POXSAT    & Pulse oximetry saturation (rest) \\
POXSAT\_WDT6 & Pulse oximetry saturation after 6-minute walk test \\
LOXAT     & Long-term oxygen supplementation \\
DIST      & 6-minute walk distance \\
BORG      & Borg dyspnea scale \\
\midrule
\multicolumn{2}{l}{\textit{Anthropometric}} \\
HEIGHT    & Height \\
WEIGHT    & Weight \\
alter     & Age \\
\midrule
\multicolumn{2}{l}{\textit{Clinical \& medication}} \\
COPDSYM   & COPD symptom score \\
PY        & Pack-years (smoking history) \\
MEDDIS    & Medication dispensation \\
\bottomrule
\end{longtable}

\end{document}